\documentclass[12pt,preprint]{aastex}

\slugcomment{Submitted to the Astrophysical Journal}
\shorttitle{Magnetic Field Comparisons in GF 9}
\shortauthors{Poidevin, F. & Bastien, P.}

\begin{document}  

\title{Comparison of Magnetic Field Structures on Different Scales in
and around the Filamentary Dark Cloud GF 9}
\author{F. Poidevin \& P. Bastien}
\affil{D\'epartement de Physique et Observatoire du Mont-M\'egantic, Universit\'e de Montr\'eal, C.P. 6128, Succ. Centre-ville, Montr\'eal, Qu\'ebec H3C 3J7, Canada}
\email{Poidevin@astro.umontreal.ca, Bastien@astro.umontreal.ca}

\begin{abstract}
New visible polarization data combined with existing IR and FIR polarization data are used 
to study how the magnetic field threading the filamentary molecular cloud GF 9 connects
to larger structures in its general environment. 
We find that when both visible and NIR polarization data 
are plotted as a function of extinction, there is no evidence for a plateau or a saturation 
effect in the polarization at $A_{\rm V} \approx$ 1.3 as seen in dark clouds in Taurus. 
This lack of saturation effect suggests that even in the denser parts of GF 9 we are still 
probing the magnetic field. 
The visible polarization is smooth and has a well-defined orientation. The
IR data are also well defined but with a different direction, and the FIR data in the
core region are well defined and with yet another direction, but are randomly
distributed in the filament region. 
On the scale of a few times the mean radial dimension of the molecular cloud, it is as if the magnetic field 
were `blind' to the spatial distribution of the filaments while on smaller scales within 
the cloud, in the core region near the IRAS point source PSC 20503+6006, polarimetry shows a rotation
of the magnetic field lines in these denser phases. 
Hence, in spite of the fact that the spatial resolution is not the same in the visible/NIR and in the FIR 
data, all the data put together indicate that the field direction changes with the spatial scale. 
Finally, the Chandrasekhar and Fermi method is used to evaluate the magnetic field strength, 
indicating that the core region is approximately magnetically critical.
A global interpretation of the results is that in the core region an original poloidal field could have 
been twisted by a rotating elongated (core$+$envelope) structure. There is no evidence for turbulence
and ambipolar diffusion does not seem to be effective at the present time.

\end{abstract}


\newpage

\section{Introduction}
        With the measurement of visible linear polarization of field stars, 
followed by developments of Near and Far Infrared
(respectively NIR and FIR) and submillimetric polarimetry, it is 
now possible to trace magnetic field
structures from the scale of the Milky Way to that of star forming
regions and cores. 
In the diffuse Interstellar Medium (ISM), the general assumption that elongated dust 
grains usually have their smallest axis aligned in a direction  
comparable to that of the local magnetic field has lead to a toroidal representation of 
the Galactic magnetic field.  
A similar topology of the magnetic field can also sometimes be observed 
in other galaxies (e.g. \citet{Scarrott1990}).
On the smaller scales of molecular clouds, a similar relation exists 
between grain alignment and magnetic field directions (see \citet{Lazarian2003}
for an exciting review on the subject) and many 
topologies of the magnetic field are suspected to exist.
Among others, helicoidal, toroidal, poloidal and hourglass configurations 
have been proposed to interpret observed polarization patterns, taking 
care of the physical conditions in these environments (e.g., Orion A,  M17). 
However, one of the most important limitations of linear polarimetry
is that we only get a two-dimensional projected average representation of the
magnetic field topology because of the integration on the plane of the sky of
the contribution to the polarized emission due to each aligned grain, 
or its dichroic absorption for visible/NIR observations. This leads to 
a degeneracy problem since many three dimensional (3D) topologies of the field can produce 
identical polarization patterns. 
A method to get rid of this degeneracy problem is the use 
of models. On the scale of relatively nearby filamentary molecular clouds 
and cores, this method has been used and allows in some cases 
to select one field topology among many possible ones 
(e.g., \citet{Matthews2001}, \citet{Vallee2003}).

        In addition to models, other diagnostics of magnetic fields can be used to lift
the degeneracy due to the plane-of-sky integration. 
When column densities are high enough, observations of the Zeeman
effect can yield the line-of-sight component of the magnetic
field strength (e.g. \citet{Crutcher1999}).  
Finally, by determining the ion-to-neutral molecular line-width ratio in non turbulent
regions with a linear flow, it is possible to determine the angle between the field
direction and the line of sight (\citet{Houde2000a}, \citet{Houde2000b}, \citet{Houde2001}). 
By combining all three methods together, we get the 3D field topology (e.g., 
\citet{Houde2002} for M17, \citet{Houde2004} for Orion A). The observation of Zeeman splitting of spectral
lines requires bright dense cores, but even in other regions, it is possible
to get the 3D geometry of the field even if the field strength is unknown.   

        These methods can inform us about how magnetic fields can contribute  
to cloud dynamics, but our ultimate goal is to understand the competition
between gravitational, thermal, magnetic, and turbulent forces. 
More specifically, magnetic fields and turbulence are both competing to slow
down the star formation process, otherwise stars would form much faster than 
observations show. Combining polarization modeling techniques (e.g., \citet{Fiege2000}) 
and predictions for magnetized simulations of turbulence
(e.g., \citet{Heitsch2001}, \citet{Padoan2001}), one can hope to be able to
determine the relative contributions of magnetic fields and turbulence. 

        To reach this goal, and to understand how magnetic fields thread the dark filaments,
and to put observational constraints on modeling 
we propose to make comparisons of magnetic fields as seen in and around individual objects.
This approach, namely a multi-scale analysis of the magnetic field,
has begun to be done via comparisons of visible, NIR, FIR and/or submm
polarization data in the OMC1 (\citet{Schleuning1998}), OMC3 (\citet{Matthews2001}) 
and GF 9 regions (\citet{Jones2003}).
In these works, aspects of the galactic magnetic field extending on tens 
of degrees are compared with aspects of the clouds probed in the densest
regions of the clouds with resolutions $\sim 10''$, but there is generally a lack 
of information between these two scales. Here we present observations 
at an intermediate scale, as much as the presence of field stars allows it, 
to have a clearer view of the topology of the field at the interface 
ISM/filamentary molecular clouds. From a general point of view, this approach
should provide new insights on the way
magnetic fields and clouds are dynamically coupled and shaped.
In this paper we address this problem for the GF 9 filamentary
molecular cloud region. A forthcoming paper will address the more complex
situation for the Orion A region. 

A description of data available about GF 9 is given in section \ref{GF9REGION}.
Observations and observational results are presented in section \ref{OBSRES}. 
An analysis and a discussion follow in section \ref{DISCUSS}, and 
our conclusions are given in section \ref{CONCLU}.

\section{The GF 9 Region} \label{GF9REGION}

The GF 9 dark filamentary structures have been catalogued first as B150 by \citet{Barnard1927},
L1082 by \citet{Lynds1962} and GF 9 by \citet{Schneider1979}. A first estimate of the distance 
to these condensations 
as $d=440 $ pc was determined by \citet{Viotti1969} (See \citet{Hilton1995}) but an estimate
of $d = 200$ pc to the extreme class 0 object GF 9-2 based on star counts was 
reported more recently by \citet{Wiesemeyer1998}. The first estimate is reported in  
catalogs or works on relatively large core samples by \citet{Benson1989}, 
\citet{Goodman1993}, \citet{Dobashi1994}, \citet{Lee1999}, \citet{Furuya2003},
while the second one is reported in works by \citet{Andre2000}, \citet{Furuya2003} and 
\citet{Froebrich2005}. Assuming  $d = 200$ pc, GF 9-2 (see Figure \ref{skymap}) has a luminosity 
$L_{\rm bol}=0.3 L_{\odot}$, an envelope mass $M_{\rm env}
 \sim 0.5 M_{\odot}$, a ratio $L_{\rm submm}/L_{\rm bol} \sim 10 \%$ and a 
bolometric temperature $T_{\rm bol} \le 20$K.
No outflow manifestation, nor structures are detected but infall motions are 
observed (\citet{Gusten1994}, \citet{Wiesemeyer1997}, \citet{Wiesemeyer1998} and 
\citet{Andre2000}).
GF 9-2 must be particularly young since it has a H$_{2}$O maser but no detected outflow 
(\citet{Furuya2003}, \citet{Furuya2005}). 

Two $\approx 8'\times10'$ regions, namely GF 9-core and GF 9-fila, have been observed by 
\citet{Ciardi1998}, \citet{Ciardi2000}. The two areas are shown in Figure \ref{skymap} by blue boxes 
identified by C and F.
The GF 9-core region is associated with the IRAS point source
class 0 protostar PSC 20503+6006, but also with GF 9-2 and an object named [LM99]351. 
JHK photometry of these two regions probe similar masses but 
the filament region is consistent with a uniform-density cylindrical cloud, 
while the core region contains a centrally condensed core with the highest extinction region 
lying to the north of the IRAS point source. CO,  $^{13}$CO and CS surveys of these 
two regions give consistent views with these two pictures, but also reveal
infall/outflow motions in order to explain CS line broadening in the core region
where n$_{\rm H_{2}} \approx 15000 \pm 3000$ cm$^{-3}$ (assuming $d =440$ pc). In the filament region, 
n$_{\rm H_{2}} \approx 6000 \pm 1200$ cm$^{-3}$ (see Table 2 in \citet{Ciardi2000}).
CO and NH$_{3}$ observations of three regions with an infrared association are presented
by \citet{Benson1989}. These sources are IRAS 20520+6003, 20526+5958 and 
20503+6006, and they are associated respectively with the LDN 1082A, B and C cores. An analysis
of these data by \citet{Goodman1993} shows that motions consistent with uniform rotation in
dense cores are present in LDN 1082A and LDN 1082B, while not in LDN 1082C.
However recent works by \citet{Furuya2005} show a velocity gradient along 
the major axis in the core in GF 9-2 which could be due to a rotation of the core.
$^{13}$CO observations by \citet{Dobashi1994} have also been done at 3 positions in 
GF 9. Referring to \citet{Lee1999}, objects named [LM99]349 and [LM99]350 (see figure \ref{skymap})
have no associated embedded young stellar objects (EYSO) while [LM99]351 ($\approx 1'$ west 
of GF 9-2) and [LM99]352 have associated EYSO.  

First $\lambda = 184$ $\mu$m FIR
polarimetry observations from space with ISO in GF 9 were done in the core and filament regions by 
\citet{Clemens1999}. Additional measurements at $\lambda = 1.65$ $\mu$m
on 9 stars shining through the 
same regions of GF 9 are also reported by \citet{Jones2003}. Finally, 850 $\mu$m polarimetry 
observations were done at the JCMT by Jane S. Greaves in 1998 at one position in the filament.

Initially, the relatively small heliocentric distance of GF 9, its filamentary 
shape, and potentially interesting FIR measurements in this
region were the major factors which motivated us to conduct
visible polarimetry around these filamentary dark structures.         

\section{Observational results} \label{OBSRES}

\subsection{Observations} \label{DATARED}

The observations were carried out on the 1.6 m telescope at the Observatoire du 
Mont-M\'egantic (OMM), Qu\'ebec, Canada,
between 2000 September and 2003 July using a 8.2 arcsecond aperture hole and a broad band red
filter (RG645: 7660 {\AA} central wavelength, 2410 {\AA} FWHM). Polarization data 
were taken with Beauty and the Beast, a two-channel photoelectric polarimeter, which 
uses a Wollaston prism, a Pockels cell, and an additional quarter wave plate. The data 
were calibrated for polarization instrumental efficiency, instrumental polarization (due to the telescope
mirrors) and zero point of position angles using a prism, non polarized and polarized 
standard stars, respectively. On average, the instrumental polarization was $0.081 \pm 
0.061 \%$ and was subtracted.
The observational errors were calculated from photon statistics 
and also include uncertainties introduced by the previously mentioned calibrations.
The final uncertainty on individual measurements of the polarization $P$ is 
usually around $0.1 \%$. For more details on the instrument and the observational 
method, see \citet{Manset1995}.   

The JCMT data archive has been used and a treatment of the data taken in 
1998 was done with commonly used SCUBA data reduction routines. 
Data were taken in relatively strong water vapour absorption ($\tau_{\rm CSO} > 0.08$) conditions,
and we finally got very low signal to noise polarization ratios and no significant results 
out of noise were found.

\subsection{Visible polarization measurements} \label{VISDATA}

Most of the stars were chosen with the GSC interface at the Canadian Astronomy
Data Center (CADC).
The others were selected directly on Digitized Sky Survey (DSS)
red plates.
All stars observed are compiled in Table \ref{PVISDATA}. 
Column 1 gives the GSC number. Column 2 shows a GF 9 star\footnote{We use an S before star numbers in 
GF 9 to distinguish them from other numbers already in use to designate sources detected at longer
wavelengths.}
 designation number used in this paper. 
Columns 3 and 4 give equatorial coordinates at epoch 2000. Degrees of polarization 
and equatorial position angles with their uncertainties are given in columns 5 and 6 respectively.
Column 7 gives the visible extinction coefficient $A_{\rm V}$ from the \citet{Dobashi2005} atlas.
Finally, 1 is tabulated in column 8 when $P > 3 \sigma_{\rm P}$
(meaning $\sigma_{\theta} < 9.5^{\circ}$), otherwise 0 is tabulated in this column.
We note that when a star has no GSC number, not much
information is generally available about it.

Figure \ref{polmap} presents a polarization map of stars for which the polarization 
degree meets the condition 
$P > 3 \sigma_{\rm P}$. This map is to be compared to Figure \ref{skymap}.
The histogram of the degree of polarization is shown in Figure \ref{phisto}. If we only consider data for 
which $P > 3 \sigma_{P}$ (data shown with full lines), the mean and the  
standard deviation of the distribution 
are $\overline{P}  = 2.22 \%$ and $S_{P} = 1.33 \%$, respectively. 
The histogram of the position angles is shown in Figure \ref{anghisto}. 
If we only consider data for which $\sigma_{\theta} < 9.5^{\circ}$ (data shown with full lines), 
the mean position angle and the standard deviation of this well peaked distribution 
are $\overline{\theta} = 127.4^{\circ}$ and $S_{\theta} = 25.5^{\circ}$, respectively.  
This means a standard deviation of 0.445 rad, comparable to those found for 
clouds with embedded clusters (see Myers $\&$ Goodman 1991b).       

Stars S35, S70, S75, S86 and S92 are identified on both maps in Figures \ref{skymap} and 2. 
Stars S35 and S86 are suspected to have variable linear polarization. 
These stars 
are located in the continuity of two filaments but in gaps where extinction 
appears to be relatively low. The data relative to these stars are compiled in Table \ref{VARDATA}
with the respective Julian Date (J.D.) of each measurement appearing in 
the last column. 

The position angle of stars S70, S75 and S92 differ significantly from the 
mean direction of the other stars ($\overline{\theta} = 127.4^{\circ}$ with $S_{\theta}= 25.5^{\circ}$). 
Star S70 is a visual binary whose components fit both in the aperture hole used for the observations;
its position angle is perpendicular to this mean position angle.
Thus, we may be in the presence of a physical binary system with circumstellar material, 
or of a single system presenting intrinsic polarization and a background star. 
The difference of the position angles of stars S75 and S92 with the average value
may be related to their position relative to the clouds as seen on Figure \ref{skymap}.  
Star S75 ($A_V = 3.3$) is located north-east of 
IRAS 20520-6003 at the edge
of two filamentary structures mainly parallel.
Comparatively, star S92 is also located
in a dense region ($A_V = 2.1$) in a small filamentary structure oriented mostly north-south. 
Given their large extinction, these two sources could either be deeply embedded in their clouds 
or be young stellar objects surrounded with disks in which multiple scattering can 
produce a relatively high degree of polarization (e.g., \citet{Bastien1990}).  
Without multiwavelength data it is difficult to distinguish between these possibilities. 

Three stars in our sample, GF 9 S1, GF 9 S2 and GF 9 S15 have measured spectral types and
visual magnitudes. However we could not get reliable distances from these values. The $A_V$
values obtained are negative or yield very large distances. Similarly using $A_V$ values
from \citet{Dobashi2005}, we could not get reliable distances. 
 
\subsection{Variations of $P$ with A$_{\rm V}$} \label{PVERSUSAV}

In order to study the variations of the degree of polarization with 
visible extinction, we used the \citet{Dobashi2005} catalog to determine
$A_{\rm V}$, in the direction of each star for which polarization measurements are available.   
This catalog is the first version of the atlas and catalog of dark clouds 
derived by using the optical database Digitized Sky Survey I (DSS), and 
applying a traditional star-count technique to 1043 plates contained in the DSS.
Using the visible extinction coefficient $A_{\rm V}$ listed in Table \ref{PVISDATA}, values of 
$P(\lambda=7660{\rm \AA})$ against $E_{\rm {B-V}}$, where $E_{\rm {B-V}} = \frac{A_{V}}{3.1}$, 
are shown in Figure \ref{pav}. For all stars, except 
GF 9 S35 and GF 9 S86 (suspected variables), data directly observed at this wavelength 
are shown with crosses. Assuming NIR data from \citet{Jones2003} and visible data probe 
the same types of grains, the Serkowski law, 

\begin{equation} \label{PAV}
P(\lambda) = P_{max} \rm {exp }[-K ln^{2}(\frac{\lambda_{\max}}{\lambda})],
\end{equation}
where $\rm{K}=1.15$, is used to derive the relation $P_{\rm V} = 1.97 P_{\rm H}$. 
This relation is used to convert the H-band data to equivalent $\lambda = 7660 {\rm \AA}$ 
data, assuming $\lambda_{\rm max} = 7660 {\rm \AA}$. 
These equivalent data are listed in Table \ref{IRTOVIS} 
and are shown with diamonds in Figure \ref{pav}.
As was previously mentionned by \citet{Jones2003}, no saturation of $P$ can be seen at about 
$A_{\rm V} \approx  1.3$ (or $E_{\rm {B-V}} \approx 0.4$), and $P_{\rm V} \approx 1.6 \%$, 
as observed in cold dark clouds 
in Taurus (See \citet{Arce1998}). Data in the visible confirm this analysis
but over the whole filament and the region surrounding it, and not just in the core region. 
Thus, while not universal,               
it is possible in some regions to probe magnetic fields inside dark clouds via polarimetry 
of  background stars as suggested here by data taken in the visible 
and in the NIR.

\section{Analysis and Discussion} \label{DISCUSS}

\subsection{Ambient Magnetic Field Orientation in the Vicinity of GF 9} \label{AMBIENTFIELD}

\begin{figure}
\plotone{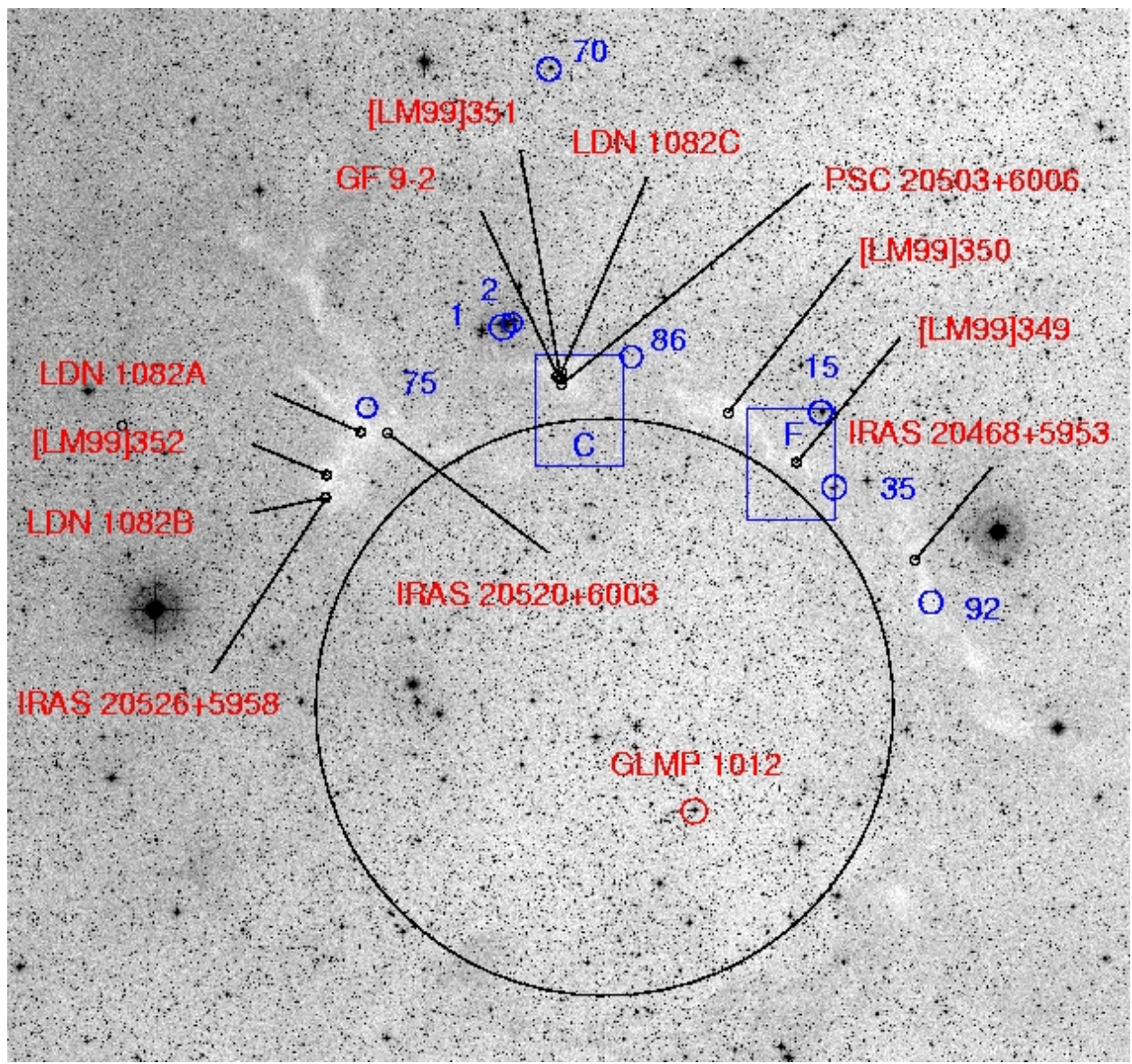}
\caption{
DSS red plate of GF 9 and its environment with reverse colors. 
Various regions appearing in 
catalogs are identified in the Figure. GLMP 1012 is a post AGB star (see text for explanations). Boxes 
are regions observed by \citet{Ciardi1998}, 
\citet{Ciardi2000} with letters C and F indicating the core region and  
the filament regions respectively. Positions of stars 
S1, S2, S15, S35, S70, S75, S86 and S92 are shown. The 
`constellation' formed by stars S75, S70, S86, S35 and S92 can be used to compare the polarization 
pattern shown in Figure \ref{polmap} with the location of the densest regions of the ISM as seen 
in the visible. The great black circle suggests 
the circular shape of the GF 9 filaments. The densest parts of the ISM forming
GF 9 are located to the north of this circle, and small faint clouds are located in the south.
\label{skymap}}
\end{figure}

As can be seen in Figure \ref{polmap} and in the histogram of position angles 
in Figure \ref{anghisto}, the polarization pattern in the vicinity of GF 9 is relatively well 
oriented on the plane of the sky with a mean direction of $\overline{\theta} = 127.4^{\circ}$ and a
standard deviation $S_{\theta} =  25.5^{\circ}$. 
Since the filament has an incurved concentric shape reminiscent of supernova remnants
(\citet{Schneider1979}), 
one can ask if a possible object at the origin of this shape could be located somewhere in the 
region delimited by the black circle traced in Figure \ref{skymap}. A possible candidate seems to be 
the post AGB star GLMP 1012 (see Figure \ref{skymap}) but given its distance estimate 
$d=$ 3 kpc (\citet{Preite-Martinez1988}) compared with those of GF 9, even by considering 
$50$ km s$^{-1}$ winds, it is difficult 
to defend the view that this source can be responsible for the shape of the filaments. 
However since the problem is tridimensional, a source at the origin of such a shape 
could also be located outside of the domain defined by this circle. Maybe such 
a source exists but has not been detected, or the apparent concentric shape
of the dense ISM is only a projection effect and other mechanisms, such as turbulent fluctuations 
in the ISM or density waves, are at the origin of its formation. A quick look in the catalogs
reveals no potential candidate in a circle a few degrees in diameter around the 
position R.A.(2000) = 20$^{\rm h}50^{\rm mn}53^{\rm s}$ and Dec.(2000) $= 59^{\circ} 51' 59''$.     
   
If now one compares the direction of the inferred magnetic field with the 
orientation of the filamentary structures on the plane of the sky, it seems 
that the magnetic field has approximately the same orientation over an arc circle
which covers about 135$^o$, from a position angle (P.A.) = 270$^o$ to 45$^o$ as
measured from the center of the circle in Figure \ref{skymap}.  
When looking at both maps from east to west in Figure \ref{skymap} and 2, a slow rotation of 
the direction of the field is present, the average directions near the core and near the
filament regions differ by about 10$^o$ (see Table \ref{MEANS}),
significantly under the variation in P.A. of the filament over the whole region. 
Moreover from east to west, there is a
great dispersion in the orientation of the filamentary structures, thus it is hard 
with this two dimensional representation to see a direct impact of the field on  
the apparent shape of the filaments.
This is in agreement with the fact that in other regions
there is no prefered projected angle on the plane of the sky 
between the directions of the ambient magnetic field and the orientation of the filamentary structures
(e.g. \citet{Myers1991b}, \citet{Heyer1987}). See also \citet{Heiles2000a} 
for a discussion on this subject.

\subsection{Multiscale Analysis of the Magnetic Field} \label{MULTI}

\subsubsection{From filamentary clouds scale to cores scale} \label{TOSMALL}

Comparisons of the orientation of the magnetic field according to the 
visible, IR and FIR data around and in the core and the filament regions can be done 
by looking at the maps in Figures \ref{mapcore} and \ref{mapfila} respectively. 
In the core region, when going from visible to IR, to FIR data, one can see 
an anticlockwise rotation of the P.A.  by about 63$^o$ in total.
For each of these two regions, 
the means and dispersions of the degree of polarization and of the P.A. 
of the visible, IR and FIR data are compiled in Table \ref{MEANS}. 
The P.A. for the FIR data corresponds to the observed polarization, rotated by 90$^o$ to
give the direction of the magnetic field. In Figures \ref{mapcore} and \ref{mapfila}, 
only FIR data for which the P.A. is approximately the same
when using the KLC instrumental polarization and the C-off instrumental polarization 
(Clemens 1999) are shown. The degree of polarization of the FIR data is not considered in the present 
analysis. Table \ref{MEANS} also includes similar information for the whole GF 9 region 
and for the local galactic scale, to be discussed below.  

\it{Comparison of Visible and IR data}
\rm

First, visible and IR data do not probe the same regions on the plane of the sky. 
Visible data probe the more diffuse parts around prestellar cores while the
IR data probe the denser phases where protostellar cores are present.
Secondly we note that in the core region the IR vectors have approximately the same orientation 
which implies that the IR polarization is not intrinsic to each star but 
produced by dichroic absorption by aligned grains. 
This statistical argument, however, does not apply in the filament region 
where only one IR measurement with sufficient signal-to-noise ratio is available.
Thirdly, in both cases the resolution used with each technique is comparable. 
Fourthly, the simple fact that the `IR' magnetic field has not the 
same orientation than the `visible' one implies that if the grains observed with 
these two techniques have identical properties, and if their dichroic absorption 
properties are the same at both IR and visible wavelengths,
then the net polarization produced by aligned grains located in the densest parts 
of the cloud dominates the net polarization produced by aligned grains located 
in the foreground diffuse ISM. 
All these facts, in addition to the dependence of $P$ with $A_{\rm V}$ mentionned in section 
\ref{PVERSUSAV}, suggest that there is a real rotation of the magnetic field lines 
in the core region when moving across the plane of the sky.

\it{Comparison of IR and FIR data}
\rm

Firstly, IR and FIR data probe the same regions on the plane of the sky
in both the core and the filament regions.
Secondly we note that in the core region the FIR vectors are very well aligned but not 
with the same P.A. than the IR vectors. IR data result from a magnetic field oriented 
at P.A. $\approx 171^{\circ}$ while FIR data suggest a magnetic field oriented at P.A. 
$\approx 19^{\circ}$.
Thirdly, the resolution with these two techniques is not the same. The FIR data resolution 
is $\approx 80''$, many times the `pencil' beam resolution of the IR and visible data.    
Finally, FIR observations probe the radiation emitted by grains mainly located in the densest 
parts of the cloud while IR observations probe grains located between the stars and 
the observer. 

When looking at Figure \ref{mapcore}, we see that IR and FIR data cover a
common area $\approx 4' \times 4' $ in size. Thus if these two
observational techniques
probe the same column densities but at two different resolutions, one
should be able to
explain the rotation of the magnetic field lines as seen at these two wavelengths.
A first hypothesis is that a given magnetic field morphology
leads to several orientations according to the resolution at which it is observed
(See \citet{Fiege2000}, but also \citet{Heitsch2001}). However, here
we see
that IR data are spatially distributed through FIR data, thus by using a
simple principle
of juxtaposition it seems impossible to reproduce the mean orientation
of FIR vectors
seen through $\approx 80''$ beams by combining several IR vectors with
very small cross
sections. On the other hand, since the mean position angle of the stars
observed in the visible
is not the same than the one for stars observed in the IR, these IR-observed stars
should not be foreground to the clouds, but rather embedded in or background to
the clouds.
In what follows we review possible explanations for the shift in position angles.
 
(1) The stars observed in the IR could be located inside the GF 9 core.
If this was the case, the impact of their presence on their immediate
environment should be observed.
IRAS observations of these stars show that their luminosities satisfy
$L_{\rm submm} / L_{\rm bol} \sim 10 \%$ (see section
\ref{GF9REGION}), which would argue against them being Class 0 sources, i.e., very
young. Therefore, we reject the hypothesis that these IR stars are all embedded in the
GF 9 core. 

(2) The stars observed in the IR could be background to the GF
9 core region.
This is suggested by their position in the $P - E_{\rm B-V}$ diagram shown
in Figure \ref{pav}.
Two possibilities can be invoked to explain the difference
in position angles:\\
(a) A first and may be naive approach is assume
that grains along the
line of sight to the GF 9 core
are all at the same temperature, and that both the IR and FIR
observational
techniques probe all grains with the same efficiency. If this is
right, the mean position angles
observed at both wavelengths should be similar, except for the known 90$^0$ difference
between them. In this scenario, the difference
in position angles is explained by a second cloud or filamentary structure located
behind the GF 9 core, and in which dust grains would be aligned differently.\\
(b) A second and probably more realistic
scenario is to emphasize differences related to the intrinsic nature of
both observational techniques. Based on point (1) mentioned above, the
assumption
of a dust temperature gradient for an externally heated cloud seems reasonable. 
In this case, the FIR measurements should be more sensitive to ``cold''
dust grains aligned in the
densest regions of the core than to ``hot'' dust grains aligned in the
envelope and in the diffuse ISM.
On the other hand, IR observations should
probe all the grains along the line of sight, independently of their temperature. 
Thus grains aligned differently in the envelope than in the denser core 
should both contribute to the IR polarization position angle,
explaining the offset in position angles.

While scenario (2b) does not reject the possibilty of a second cloud or 
filament evoked in (2a), it has the benefit to propose a
simple and consistent explanation.
Thus we prefer scenario (2b), in which the intermediate position angles of IR vectors
shown in figure \ref{mapcore}
may result from a vectorial addition of
both FIR and visible vectors covering this region and its vicinity.

\it{Comparisons of the magnetic fields orientations with the cores elongations} 
\rm

\begin{figure}
\epsscale{0.8}
\plotone{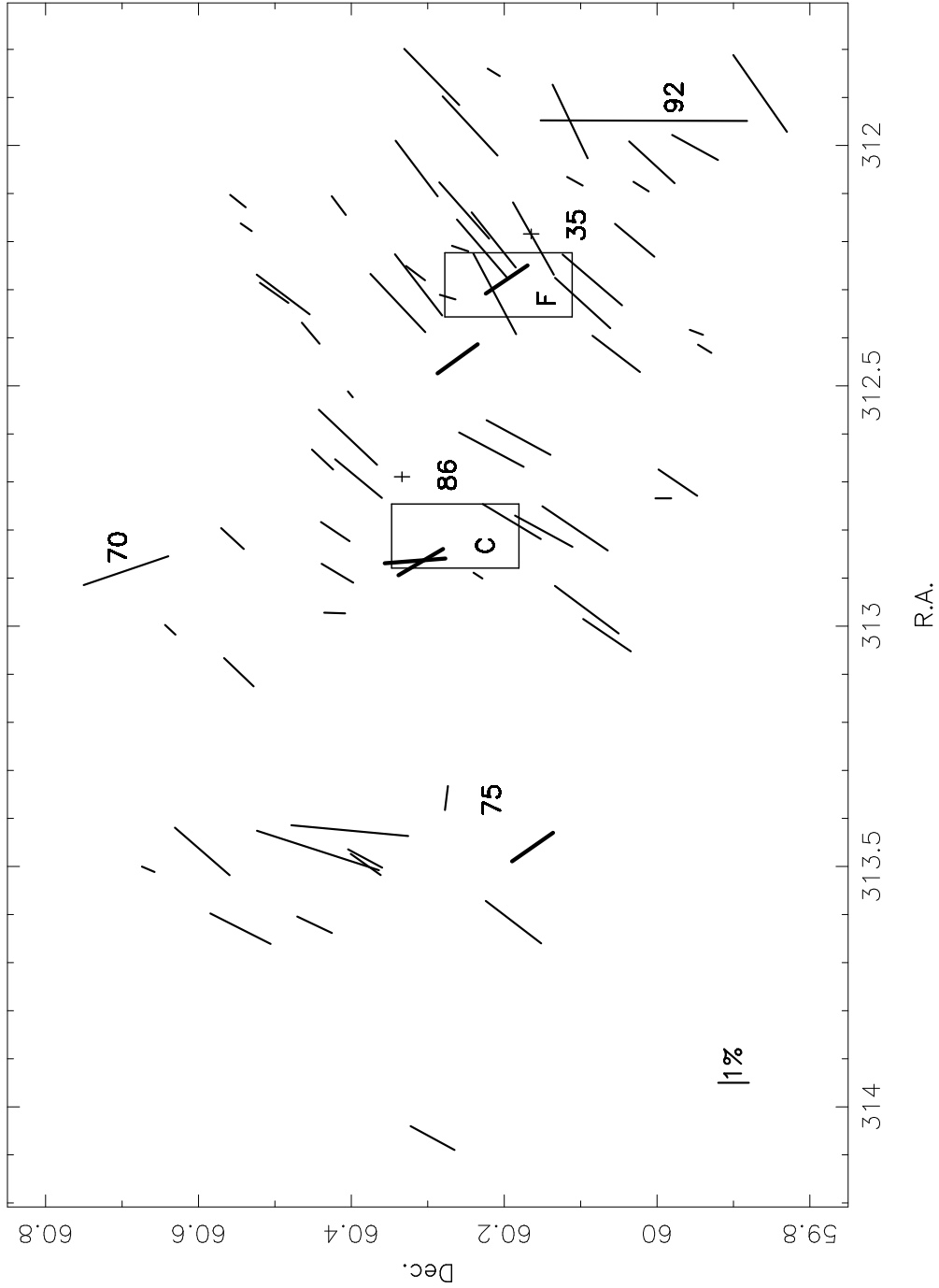}
\caption{
Visible polarization map of GF 9. The C and F rectangles are the 
core and the filament regions observed by \citet{Ciardi1998}, \citet{Ciardi2000}. 
Stars S75 and S92 are embedded in the filament 
and have a polarization with a different orientation from the mean orientation of the 
ambient magnetic field. Object
S70 is a visual binary. Crosses show the positions of stars S35 and S86 for which 
variable polarization is suspected (See Table \ref{VARDATA}). Only data for which $P > 3 \sigma_{P} $
are represented in the Figure. Bold vectors with sizes of $2 \%$
show position angles of cores LM349, LM350, LM 351, L1082C and L1082B. (See Figure \ref{skymap}).
\label{polmap}}
\end{figure}

\begin{figure}
\epsscale{0.8}
\plotone{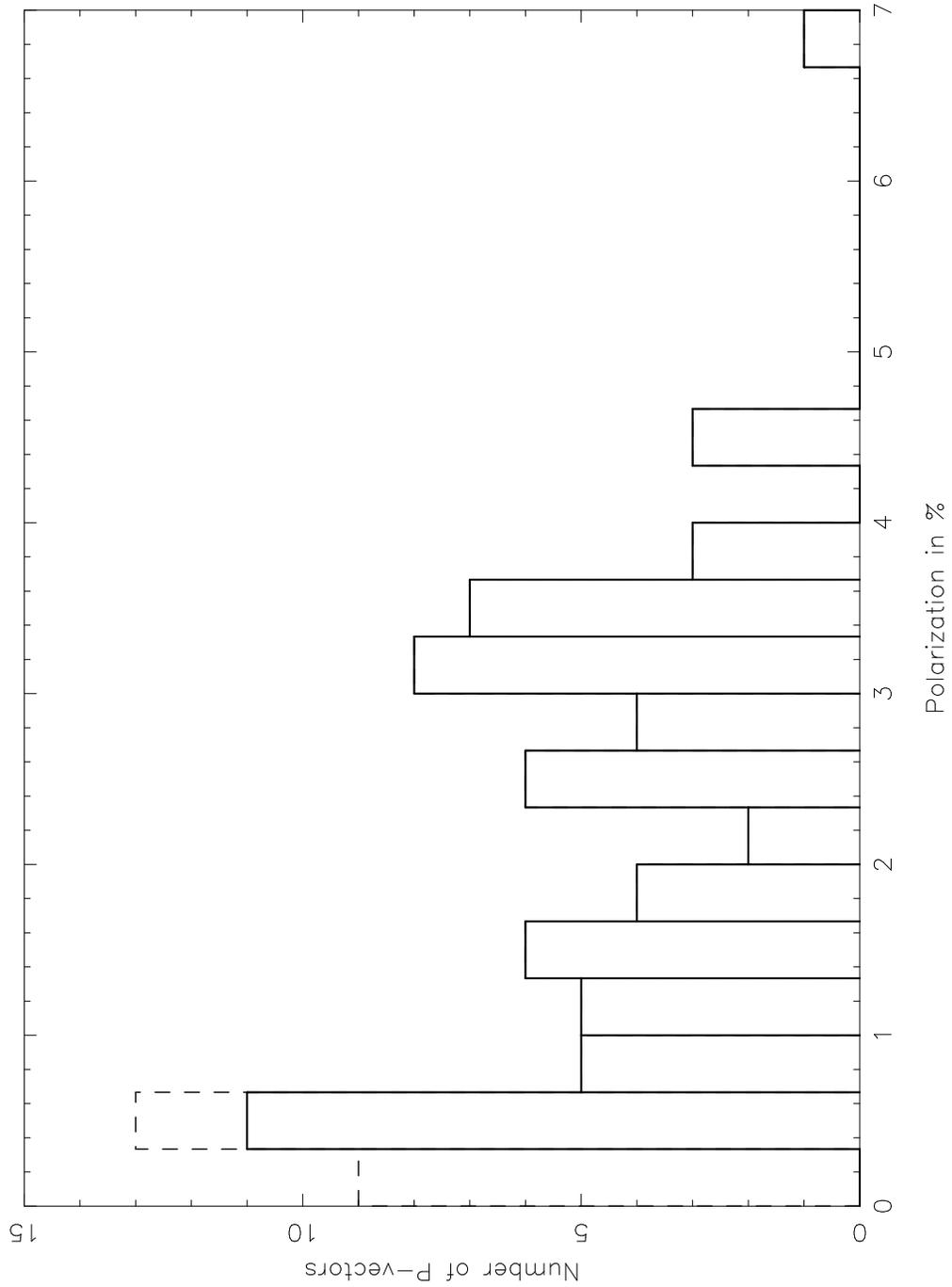}
\caption{ 
Histogram of polarization in GF 9. Data for which $P > 3 \sigma_{P} $ (See last column 
in Table \ref{PVISDATA}) are shown with bold lines. 
For this set of data, the mean and the dispersion of the distribution of the degrees of 
polarization are 
$\overline{P} = 2.22\% $ and $S_{\rm P} = 1.33 \%$ respectively. 
\label{phisto}}
\end{figure}

\begin{figure}
 \epsscale{0.8}
\plotone{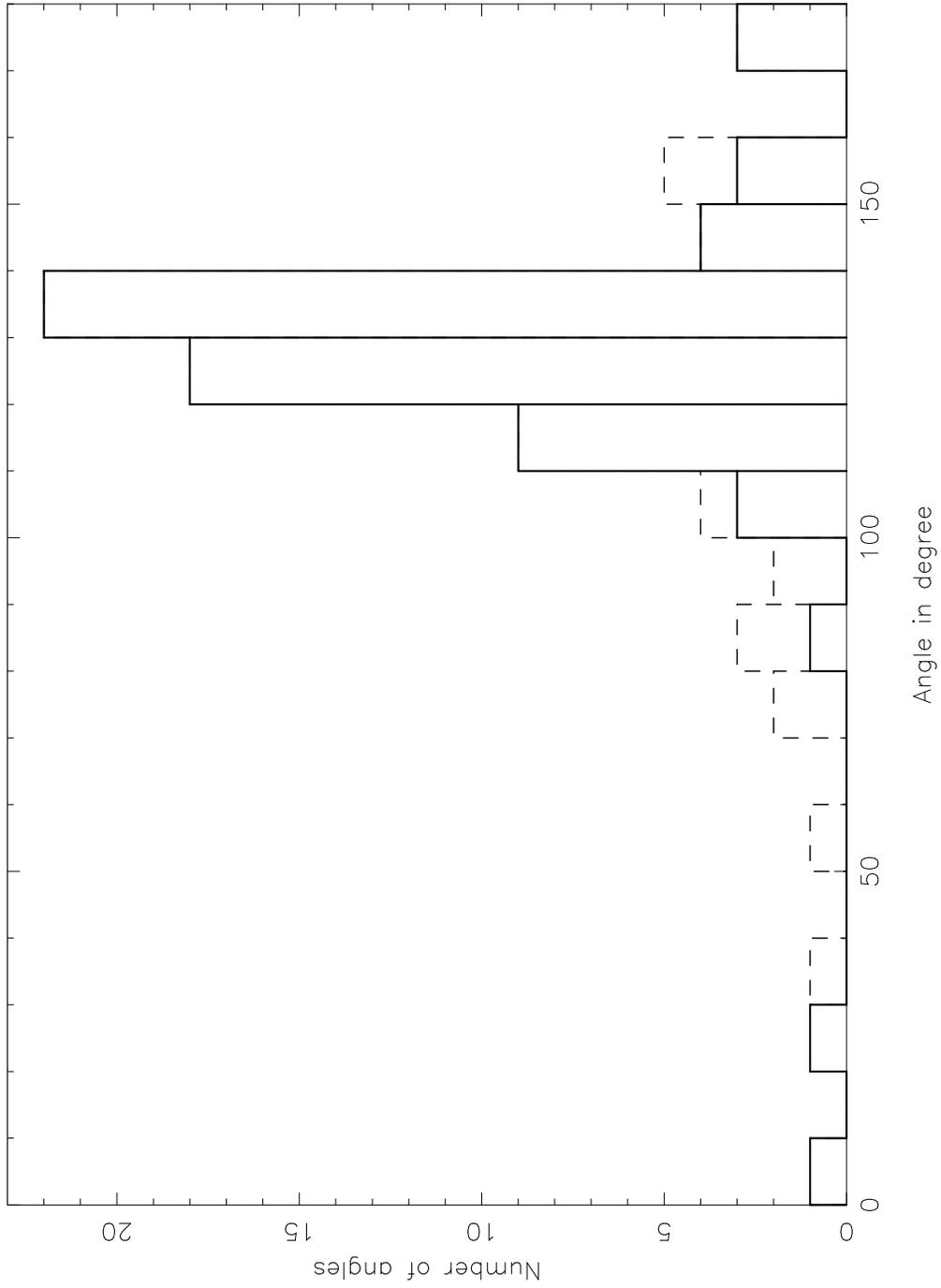}
\caption{ 
Histogram of position angles of polarization in GF 9. Data for which $P > 3 \sigma_{P} $
(meaning $\sigma_{\theta} < 9.5^{\circ}$) are shown with bold lines. 
For this set of data, the mean and the dispersion of the distribution 
of the position angles are $\overline{\theta} = 127.4^{\circ} $ and $S_{\theta} = 25.5^{\circ}$ respectively. 
\label{anghisto}}
\end{figure}

\begin{figure}
\epsscale{1.0}
\plotone{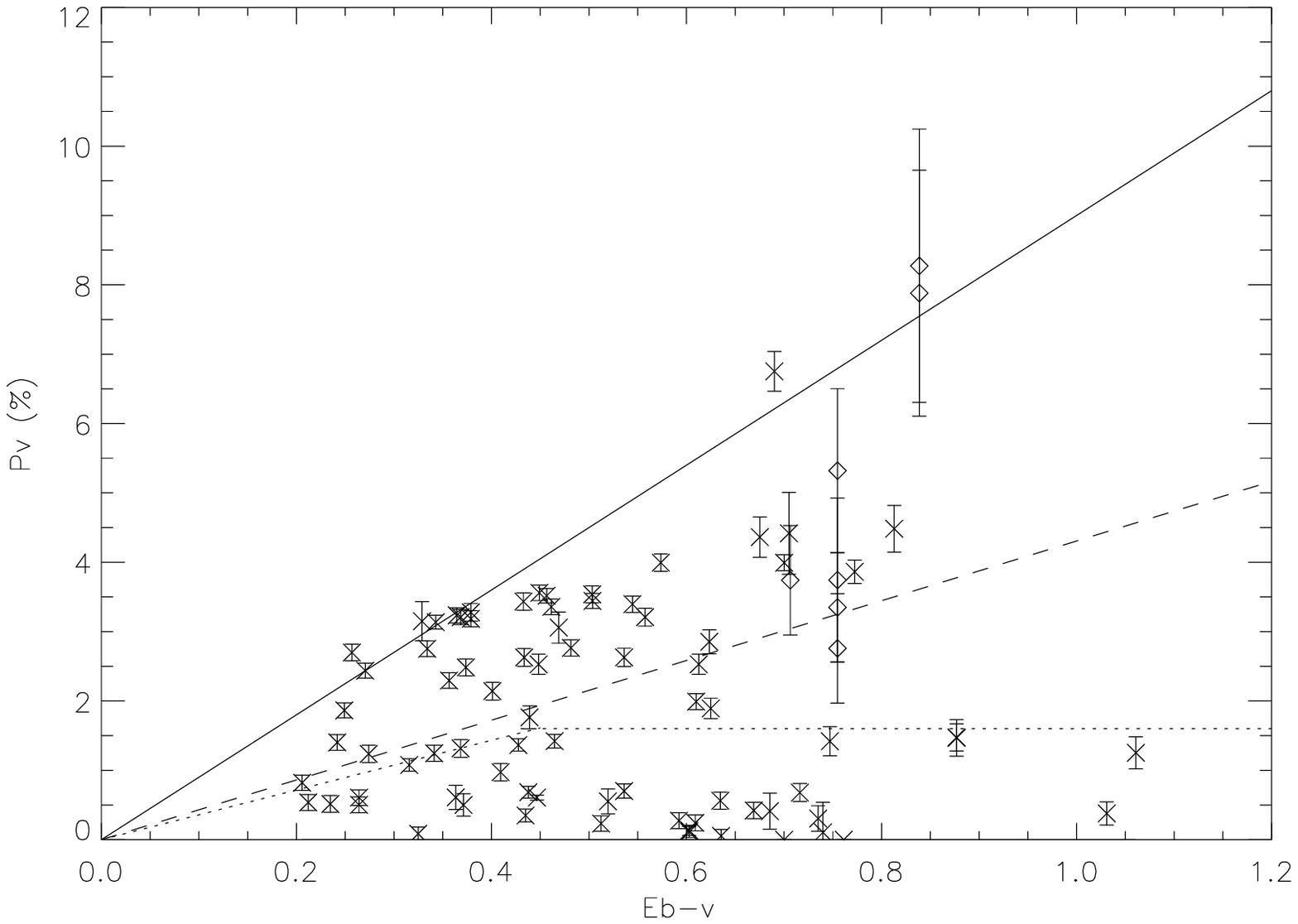}
\caption{ 
Variations of $P$ with $E_{\rm{B-V}}$.
$A_{\rm V}$ data from \citet{Dobashi2005} were used to compute $E_{\rm{B-V}}$. 
The full line delineates the 
upper envelope $P \le 9.0 E_{\rm{B-V}}$ (\citet{Serkowski1975}). A linear fit to the
whole data set is shown by the dashed line and is given by $P \approx 4.3 E_{\rm{B-V}}$. 
Dotted lines reproduce the 
truncation in the $(P, E_{\rm{B-V}})$ relation observed by \citet{Arce1998} in 
dark clouds in Taurus. X symbols: broadband red data from Observatoire du Mont-M\'egantic. Diamonds:
NIR data from \citet{Jones2003} converted to red-equivalent data (see section \ref{PVERSUSAV}).
\label{pav}}
\end{figure}

We compare in Table \ref{OFFSETS} the orientations of five protostellar cores spread
into the filaments with the magnetic field orientations. 
Orientations of the cores L1082A, B and C were determined with the $(J,K) = (1,1)$ 
lines of NH$_{3}$ while orientations of the LM cores are based on the visible DSS data
\footnote{In fact, objects L1082C and LM351 may be the same source. 
The difference in coordinates and in P. A. comes about because of the different 
observational techniques used, NH$_{3}$ for L1082C and visible extinction for LM351.}
At first sight we see that, except for L1082B, these dark condensations 
have a tendency to be elongated 
along the filaments as was noticed by \citet{Myers1991a} (see Figures \ref{skymap} and \ref{polmap}).
Names, positions and P.A. of these cores are shown in Table \ref{OFFSETS}
in columns 1 through 4 respectively. For objects LM351 and L1082C, located in the core region, 
and for objects LM349 and LM350, located in the filament region, the offset, $|\Delta \theta_{7660}|$, 
on the plane of the sky between the P.A. of each 
protostellar core and the mean magnetic field orientation probed 
with visible data is shown in column 5. The same is done with the IR and FIR measurements
in columns 6 and 7 respectively. These offsets were not estimated for L1082B since 
the magnetic field direction is undersampled around this object.

On larger scales, the magnetic field is mostly along the minor axis of the
cores, as can be seen in Figure \ref{polmap}.
However, on smaller scales, if we accept that IR measurements can probe 
the magnetic field in dense regions as it is suggested by the variations of $P$ 
with $A_{V}$ shown in Figure \ref{pav}, we see that a rotation of the field is apparent 
in the core region. On this smaller spatial scale the field is well aligned in the core
region in a direction mainly parallel to the cores' major axes.
Turbulence models such as those developed by \citet{Padoan2001}
show that matter may flow onto cores along filaments with the field 
being stretched along the filaments. In such a case, the field would tend
to be along the major axis of the elongated low-density structures connected 
with higher density cores. However, such models show that the polarization pattern probing 
dust in the low density structures would be chaotic rather than smooth and well aligned.
Moreover, \citet{Furuya2005} showed recently that a velocity gradient 
of 2.3 km s$^{-1}$ pc$^{-1}$ is present along the major axis of the core GF9-2
but none along its minor axis. 
Therefore, if we assume that sources named LM351, L1082C and GF 9-2 are associated to 
a common spatial volume traced with different tracers, and
if we assume that the velocity gradient along the major axis of GF 9-2 is due to 
rotation of the core, this picture suggests that an original poloidal field 
has been twisted on small scales by the rotation of the core into a toroidal field.  
This would be consistent with a magnetic support model.

On the other hand, this effect does not seem to be present in the filament region
where the magnetic field probed with IR data is mainly parallel to the small axes of the cores
detected in this area, in approximately the same direction than the magnetic field
seen at greater scales. But on the other hand if FIR measurements are reliable, the chaotic
FIR polarization pattern suggests that turbulence plays an active role in the filament region.    

Finally comparison of both regions suggests a magnetic field interacting differently, and maybe less 
with its surrounding medium in the filament than it does in the core region, in agreement
with the evolutionary stage of these two regions. Additional IR data would be necessary to confirm 
this point, and the quality of FIR measurements should of course be confirmed in both regions. 

\subsubsection{From the filamentary clouds scale to the local Galactic scale} \label{TOLARGE}
 
The mean orientation of the ambient magnetic field can also be compared to the local 
Galactic magnetic field orientation. 
Figure \ref{mapgal} shows a polarization map in Galactic coordinates 
based on measurements compiled by \citet{Heiles2000b} with data for 
which $P/\sigma_{P}>3$. The region presents a complicated picture because we are looking 
down a spiral arm and possibly because of the influence of the Cygnus star-forming complex.
In their works \citet{Heiles1996} and \citet{Walawender2001} have used a large set of 
polarization data and studied the inferred magnetic field geometry in this part of 
the Galaxy. At first sight, the simple picture of a toroidal magnetic field implies 
that the magnetic field lines should point in the direction toward $l= 90^{\circ}, b= 0^{\circ}$. 
In their statistical study \citet{Walawender2001} show that this trend is effectively real  
and are also able to detect the curvature of the local spiral arm. 
As mentionned by \citet{Jones2003}, since GF 9 is at a relatively close distance
it is probably threaded by the magnetic field in the local spiral arm.
Jones compares the orientation of the mean IR vector in the core region   
with the visible vectors on a greater scale, and shows that this vector points in the direction of the  
`vanishing point' approximately located at $l= 90^{\circ}, b= 0^{\circ}$.
We can see this vector denoted by `IR' in Figure \ref{mapgal}. We also show the mean visible 
vector `V' in this map. It does not point in the same direction 
than the `IR' vector but it 
is approximately parallel to its nearest neighbours. All these facts suggest that 
it is preferable to use data probing the diffuse parts of the ISM in the environments 
of molecular clouds to compare ambient to the cloud and Galactic magnetic 
field's orientations.

\subsection{Magnetic field strength and magnetic flux in the core region} \label{STRENGTH}

In order to estimate if the core region observed in CS by \citet{Ciardi2000}
is magnetically supercritical or magnetically subcritical, 
we use the \citet{Chandrasekhar1953} method to estimate the magnetic field 
strength. The magnetic field strength in the plane-of-sky
is estimated with the expression:
\begin{equation} \label{BPOS}
B_{\rm pos} \approx 9.3 \sqrt{n(\rm H_{2})} \frac{\Delta \rm (V)}{\delta \phi} \mu{\rm G},
\end{equation} 
where $\delta \phi$ is the polarization P.A. dispersion 
in degrees, $n(\rm H_{2})$ is the molecular hydrogen density in molecules cm$^{-3}$,  
$\Delta \rm V= \sqrt{8ln2}\delta \rm(V)$ is the FWHM line width in km s$^{-1}$, 
and $\delta \rm V$ is the velocity dispersion (see \citet{Crutcher2004}). 

We use only the IR data for the core region since we have no way of estimating the
turbulent velocity dispersion for the larger area corresponding to the visible data.
The value $\delta \phi = 5.9^{\circ}$ is taken from Table \ref{MEANS} using the Jones IR data. 
This value is provided by a relatively small set of six measurements for which 
the mean of the uncertainties is $6.7^{\circ}$. 
If instead of a direct average, we use a weighted average for estimating the dispersion
in P.A., we get $7.2^{\circ}$. Finally, working with the P.A. directly
instead of the vectors, the dispersions are $6.4^{\circ}$ and $6.0^{\circ}$, weighted 
and unweighted respectively.  
All these values are close to each other and reflect the fact that there is little dispersion
in P.A. in the cloud. Therefore, we believe that despite the small number of
measurements, we nevertheless have a representative value for the dispersion. 
The various parameters that we used for the CS 
core region and our results are shown in Table \ref{STRENGTHS}. 

Once $B_{\rm pos}$ is derived, we estimate $\lambda_{\rm obs}$, the observed mass to 
flux ratio $(\frac{M}{\phi_{B}})_{observed}$ normalized to its critical value.
This observed normalized mass to flux ratio is subject to geometrical bias,
as explained by \citet{Crutcher2004}. In the case of estimating the total field
from the plane-of-sky component of the field,
two of the three components are known and the correction factor for $\lambda_{\rm obs}$ 
is then $\pi/4$, which corresponds to $\lambda_{\rm c}$ given in Table \ref{STRENGTHS}.
This correction is statistical in nature since we do not know the appropriate angle for any
individual object. In fact, there may not be a line of sight component to the field. 
Using $\delta \phi = 8.9^{\circ}$ for the dispersion in P.A.
\footnote{$8.9^{\circ}$ is the quadratic sum of the dispersion $5.9^{\circ}$ and the
mean uncertainty $6.7^{\circ}$}
to take into account the 
uncertainties in the measurements would decrease the magnitude of $B_{\rm pos}$ and
increase the magnitude of $\lambda_{\rm obs}$ and $\lambda_c$ by about 50\%.
For comparison, the dispersion in P.A. for the visible data around the core region
is $12.4^{\circ}$ (Table \ref{MEANS}). With this value (even though we are outside the area for which the CS
line-widths have been measured), the values of $\lambda_{\rm obs}$ and $\lambda_{\rm c}$ 
are larger by about 110\%. These values are probably too large because the visible
data cover a much larger area than just the core region. 
Also, a bias exists in the Chandrasekhar-Fermi 
method that tends to cause estimated field strentghs to be too large and thus 
to underestimate values of $\lambda_{\rm obs}$ (e.g. \citet{Heitsch2001}). 
 
What is the effect of the distance $d$ on these results? The column density $N$ is used to get the
gas mass, so $M \propto r^2$, where $r$ is the radius of the cloud, assumed to be spherical.
The density $n \propto M/volume \propto r^{-1}$. Since $r \propto d^{-1}$,
$n \propto d$, $B_{pos} \propto d^{1/2}$ and $\lambda \propto d^{-1/2}$. If the distance is 200 pc
instead of 440 pc as assumed for the values given in Table \ref{STRENGTHS}, $\lambda$ would
be closer to unity by about $\sqrt{2}$.

Is there additional possible evidence for a subcritical core? \citet{Crutcher2004} 
identified two other tests
to distinguish between ambipolar diffusion and turbulence as supporting mechanisms for star
formation. If ambipolar diffusion is acting, one would expect an
hourglass configuration for the magnetic field. But the distribution of the IR polarization
vectors (see Fig \ref{mapcore}) suggests a smooth field, or even a slight curvature in the opposite
direction than expected for an hourglass shape. A third test identified by \citet{Crutcher2004}, 
if ambipolar diffusion is acting one expects
$B \propto \rho^{\kappa}$ with $\kappa$ approaching 1/2 for larger densitites
($\gtrsim 100$ cm$^{-3}$).
The values derived here are more or less compatible with this relation 
(see Figure 1 in \citet{Crutcher1999}).
In addition, the comparison of the magnetic field orientations with the elongations of the cores
seen in Table \ref{OFFSETS} is not in general compatible with ambipolar diffusion, since in that case the
cores would be perpendicular to the direction of the magnetic field.

Taking all the above arguments into account, our data suggest that the GF 9 core region is 
approximately magnetically critical.
Ambipolar diffusion does not appear to be supported. Also, turbulence would produce a 
randomness in the polarization vectors which is not observed. The picture presented 
in section \ref{TOSMALL} of an
original poloidal field which is transformed into a toroidal field by rotation appears most plausible.

\section{Conclusions} \label{CONCLU}

Visible linear polarization data taken at the Mont-M\'egantic Observatory 
on 78 stars in the vicinity of the dark filamentary cloud GF 9 were presented and 
compared with available NIR and FIR polarization data covering part of the same region. 
Our main conclusions are:

On a scale of a few 
times the mean radial dimension of the molecular clouds, the magnetic field is smooth and well-ordered, 
and appears to be `blind' to the spatial distribution of the filaments.  

On smaller scales in the core region, NIR polarimetry shows a rotation 
of the magnetic field lines when reaching into the denser parts.

There is no evidence for saturation of polarization as a function of extinction for both the
visible and NIR data, suggesting that even in the denser regions the same grain properties and
grain alignment mechanisms apply, and that the magnetic field is still being probed adequately there.

Finally, the magnetic field strength was estimated with the \citet{Chandrasekhar1953} method.
The results would suggest that the core region is approximately magnetically critical. 

All these facts put together suggest that {\it in GF 9 the magnetic field changes direction
with the spatial scale.} The field differs by $35^{\circ}$ in the GF 9 core compared to the surrounding 
region. Ambipolar diffusion does not appear to be supported given the current
evidence. The effects expected from turbulence are definitely missing.  
A global interpretation of the results is that an original poloidal field could have 
been twisted into a toroidal configuration by a rotating elongated structure during its collapse. 

Future work should check if the trend in polarization position angles to rotate 
counterclockwise with increasing wavelength still continues in the submm. Also, as the lack
of saturation of polarization as a function of extinction suggests, 
if dust grains are magnetically aligned there should be no
polarization hole in the submm in the core and the filament regions. As an additional
benefit, submm polarization would afford a higher spatial resolution than the FIR data discussed here. 
  
We thank the Conseil de recherche en sciences naturelles et en g\'enie du Canada
for supporting this research. 
The authors thank J. S. Greaves for providing information about 
observations of GF 9 with SCUBAPOL at the JCMT, as well as 
B. Malenfant and G. Turcotte for their helpful and friendly support during observations at 
Mont-M\'egantic Observatory. 
We also thank an anonymous referee for his constructive comments and helpful suggestions.
This work made extensive use of the SIMBAD database at the Canadian Astronomy Data Center, 
which is operated by the Dominion Astrophysical
Observatory for the National Research Council's Herzberg Institute of Astrophysics.

\clearpage
\begin{deluxetable}{lrrrrrrrrcc} 
\tablewidth{0pt}
\tabletypesize{\scriptsize}
\tablecaption{Visible polarization data. \label{PVISDATA}}
\tablehead{
  \colhead{GSC}                              & \colhead{GF9}                           &
  \colhead{$\alpha (2000)$}                  & \colhead{$\delta (2000)$}               & 
  \colhead{ $P \pm \sigma_ P$}               & \colhead{$\theta \pm \sigma_{\theta}$}  &
  \colhead{ A$_{\rm V}$}                     & \colhead{$\frac{P}{\sigma_{P}}>3$} \\
  \colhead{number}                             & \colhead{number}                          &
  \colhead{($^{\rm h}$ $^{\rm mn}$ $^{\rm s}$)}  & \colhead{($^{\circ}$ ' '')}                 & 
  \colhead{ ($\%$)}                          & \colhead{($^{\circ}$)     }                 &
  \colhead{ (Mag.)}                     & \colhead{ \tablenotemark{(a)}}}
\startdata
no&91&20:47:23.5&+60:12:59.5&0.55$\pm$0.18&136.1$\pm$9.4&1.61&1\\
0424600447  &40& 20:47:25.8  &+60:17:41.2&3.43$\pm$0.13&121.6$\pm$1.0&1.34&1\\
no&120&20:47:34.0&+59:51:54.4&4.36$\pm$0.29&113.7$\pm$1.9&2.09&1\\
0424601275&46&20:47:35.4&+60:03:35.1&0.24$\pm$0.12&87.1$\pm$13.8&1.89&0\\
no&92&20:47:47.6&+60:01:02.2&6.75$\pm$0.29&179.8$\pm$1.2&2.14&1\\
0424601151&57&20:47:48.0&+60:06:49.4&3.99$\pm$0.13&106.7$\pm$0.9&1.78&1\\
0424600411&31&20:47:50.2&+60:14:40.6&3.56$\pm$0.11&120.2$\pm$0.9&1.39&1\\
0396300254&48&20:48:01.0&+59:57:01.1&1.99$\pm$0.12&139.0$\pm$1.7&1.89&1\\
no&98&20:48:01.1&+60:00:47.1&0.41$\pm$0.26&79.9$\pm$18.1&2.13&0\\
0424600967  &47&20:48:08.4&+60:00:24.6 &2.63$\pm$0.13&124.4$\pm$1.4&1.66&1\\
0424600359   &39&20:48:11.5 &+60:18:52.2&3.21$\pm$0.11&115.4$\pm$1.0&1.14&1\\
0424601339  &36&20:48:17.9  &+60:06:26.9 &0.68$\pm$0.13&138.2$\pm$5.5&2.22&1\\
0424600585 & 18&20:48:20.5 &+60:01:15.2& 0.70$\pm$0.10  & 135.3$\pm$4.1&1.66&1\\
0424600629& 17&20:48:25.5 &+60:10:11.9& 0.06$\pm$0.09  & 37.3$\pm$49.6&1.97&0\\
0424600469 &11 &20:48:27.7 &+60:32:54.5& 0.82$\pm$0.11  &128.1$\pm$3.9&0.64&1\\
0424601143&13 &20:48:30.0 &+60:24:59.4& 1.08$\pm$0.09  & 115.2$\pm$2.4&0.98&1\\
0424601171 & 16&20:48:32.5 &+60:15:08.4& 3.36$\pm$0.11  & 119.1$\pm$1.0&1.43&1\\
0424600521 &10 &20:48:40.8 &+60:32:15.1& 0.54$\pm$0.12 & 132.5$\pm$6.1&0.66&1\\
0424600727  &35&20:48:44.2 &+60:09:52.9 &see Table \ref{VARDATA}&&2.17&\\
0424601053  &34&20:48:46.5  &+60:09:42.4 &3.99$\pm$0.12&109.5$\pm$0.8&2.17&1\\
0424600659&54&20:48:47.1&+60:12:49.4&3.21$\pm$0.13&116.7$\pm$1.1&1.73&1\\
0424600961&37&20:48:47.3&+60:01:45.4 &2.14$\pm$0.13&127.0$\pm$1.7&1.24&1\\
0424601093&53&20:48:51.3&+60:13:46.1&3.39$\pm$0.12&118.5$\pm$1.0&1.69&1\\
0424600621 &15 &20:48:51.4&+60:15:27.6& 0.61$\pm$0.04  & 151.5$\pm$1.8&1.38&1\\
0424600109 &14 &20:49:03.7 &+60:18:57.7& 0.97$\pm$0.13  & 129.9$\pm$3.7&1.27&1\\
0424601169&52&20:49:07.1&+60:05:05.0&3.28$\pm$0.13&126.3$\pm$1.1&1.17&1\\
0424600721&55&20:49:09.6&+60:18:43.4&3.54$\pm$0.13&115.8$\pm$1.0&1.56&1\\
0424600857 &9 &20:49:13.6 &+60:30:03.6& $1.40\pm$0.11  & 131.6$\pm$2.3&0.75&1\\
no&90&20:49:14.1&+60:12:44.5&4.42$\pm$0.59&108.4$\pm$3.8&2.19&1\\
0424600123& 8&20:49:14.4 &+60:29:21.0& 2.70$\pm$ 0.12 &130.0$\pm$1.3&0.80&1\\
0424601087  &38&20:49:15.7&+60:16:26.1 &0.56$\pm$0.13&155.2$\pm$6.3&1.97&1\\
0424600963&51&20:49:18.6&+60:05:50.6&3.18$\pm$0.12&124.6$\pm$1.1&1.17&1\\
0424600281&56&20:49:18.7&+60:20:20.9&3.52$\pm$0.10&120.6$\pm$0.8&1.42&1\\
0396300376 & 19&20:49:33.3 &+59:56:54.5& 0.50$\pm$0.11  & 149.4$\pm$6.4&0.82&1\\
0424600981 &7 &20:49:33.7  &+60:27:11.5 &1.24$\pm$0.13  &117.9$\pm$2.9&0.85&1\\
0396300020 & 20&20:49:41.5 &+59:56:15.7& 0.61$\pm$0.11  &135.6$\pm$5.3&0.82&1\\
0424601133& 21&20:49:44.1 &+60:03:12.3& 2.44$\pm$0.11  & 129.6$\pm$1.3&0.84&1\\
0424601015 &6 &20:50:04.3 &+60:24:04.9&0.35$\pm$0.09 & 117.0$\pm$7.6&1.35&1\\
0424601131   &42&20:50:25.7&+60:24:15.8 &3.44$\pm$0.11&123.6$\pm$0.9&1.56&1\\
0424601183&50&20:50:25.8&+60:10:52.6&2.75$\pm$0.11&139.3$\pm$1.2&1.04&1\\
0424601341   &33&20:50:31.9&+60:13:00.4 &2.77$\pm$0.12&139.7$\pm$1.2&1.49&1\\
0424601127  &44&20:50:36.8&+60:26:15.6 &1.25$\pm$0.12&123.6$\pm$2.7&1.06&1\\
0424601041   &32&20:50:37.2 &+60:15:38.6 &0.27$\pm$0.12&156.0$\pm$12.2&1.84&0\\
no&86&20:50:45.3&+60:20:01.4&see Table \ref{VARDATA}&&2.36&\\
0424600477  &41&20:50:46.3&+60:23:26.3&2.53$\pm$0.15&127.3$\pm$1.7&1.39&1\\
0396300338 & 22&20:50:48.3 &+59:58:22.0& 1.86$\pm$0.11  & 132.8$\pm$1.7&0.77&1\\
0396300620 & 23&20:50:56.2 &+59:59:30.6& 0.51$\pm$0.12  & 180$\pm$6.7&0.73&1\\
0424600655&29&20:51:07.8&+60:11:24.2&2.63$\pm$0.13&136.2$\pm$1.4&1.34&1\\
0424601145&27&20:51:11.2&+60:06:26.2&3.14$\pm$0.10&133.0$\pm$0.9&1.06&1\\
0424601213&28&20:51:12.6&+60:08:53.9&2.48$\pm$0.12&139.2$\pm$1.4&1.16&1\\
0424601111& 4&20:51:12.8 &+60:25:15.5& 1.36$\pm$0.09 &133.3$\pm$1.8&1.33&1\\
0424600103 &5 &20:51:16.2 &+60:33:19.9& 1.32$\pm$0.13&124.3$\pm$2.8&1.14&1\\
no&70&20:51:32.3&+60:41:41.3&3.15$\pm$0.28&28.4$\pm$2.6&1.02&1\\
0424701143 &43&20:51:33.6&+60:25:05.5&1.42$\pm$0.10&136.4$\pm$2.0&1.44&1\\
0424701033&45&20:51:34.7&+60:14:03.5  &0.42$\pm$0.12&133.3$\pm$8.1&2.07&1\\
0424701049&30&20:51:43.7&+60:11:50.0&0.23$\pm$0.11&52.2$\pm$13.9&1.59&0\\
0396300164& 24&20:51:49.2 &+59:54:15.5& 0.09$\pm$0.10  & 73.0$\pm$32.4&1.01&0\\
0424701155&26&20:51:51.7&+60:05:31.0&3.23$\pm$0.11&130.1$\pm$1.0&1.13&1\\
0424701211 &3 &20:51:53.4 &+60:25:18.5&0.68$\pm$0.09  & 176.8$\pm$3.5&1.36&1\\
no&85&20:51:55.4&+60:19:00.0&0.11$\pm$0.43&91.8$\pm$57.7&2.29&0\\
0424701127 &2 &20:51:57.1 &+60:22:44.6&0.13$\pm$0.07  &87.4$\pm$15.7&1.87&0\\
no&71&20:52:01.8&+60:38:13.2&0.61$\pm$0.17&123.8$\pm$8.2&1.13&1\\
0424701215 &1 & 20:52:02.7&+60:22:27.9&0.11$\pm$0.08 & 90.7$\pm$20.9&1.87&0\\
0424701115&25&20:52:04.5&+60:03:55.6&2.29$\pm$0.11&132.7$\pm$1.4&1.11&1\\
0424701085&72&20:52:23.2&+60:32:48.4&1.76$\pm$0.16&123.2$\pm$2.7&1.36&1\\
no&75&20:53:25.7&+60:16:32.6&1.25$\pm$0.23&85.7$\pm$5.3&3.29&1\\
no&68&20:53:42.0&+60:24:07.8&3.86$\pm$0.17&171.6$\pm$1.3&2.39&1\\
no&66&20:53:52.0&+60:26:37.7&4.48$\pm$0.34&152.8$\pm$2.1&2.52&1\\
no&59&20:53:52.8&+60:35:44.6&3.06$\pm$0.22&125.9$\pm$2.1&1.45&1\\
no&74&20:53:55.6&60:15:51.6&0.38$\pm$0.17&158.9$\pm$12.6&3.20&0\\
no&69&20:53:55.7&+60:22:55.9&1.47$\pm$0.20&131.6$\pm$3.8&2.72&1\\
no&69bis&20:53:59.0&+60:22:50.0&1.47$\pm$0.26&139.4$\pm$5.1&2.72&1\\
no&61&20:54:01.3&+60:39:55.3&0.50$\pm$0.16&146$\pm$9.2&1.15&1\\
no&77&20:54:27.8&+60:11:16.5&2.85$\pm$0.17&129.4$\pm$1.7&1.93&1\\
no&65&20:54:28.8&+60:26:53.5&1.42$\pm$0.21&143$\pm$4.2&2.32&1\\
0424700062&62&20:54:31.3&+60:32:40.7&2.53$\pm$0.15&141.3$\pm$1.7&1.90&1\\
no&64&20:54:53.0&+60:29:35.7&0.30$\pm$0.18&102.7$\pm$17.2&2.28&0\\
0424700686&73&20:56:15.6&+60:17:41.9&1.89$\pm$0.15&139.3$\pm$2.2&1.94&1\\
\enddata
\tablenotetext{(a)}{1 is tabulated
when $P > 3 \sigma_P$, otherwise 0 is tabulated.}
\end{deluxetable}

\clearpage

\begin{deluxetable}{lrrrr} 
\tablewidth{0pt}
\tabletypesize{\scriptsize}
\tablecaption{Stars with suspected variable polarization. 
Coordinates are given in Table \ref{PVISDATA}. \label{VARDATA}}
\tablehead{
  \colhead{GF9}                       & \colhead{$P \pm \sigma_P$}              &
  \colhead{$\theta \pm \sigma_{\theta}$}     & \colhead{J.D.}                        \\
 \colhead{number}                       & \colhead{$(\%)$}              &
  \colhead{$^{(o)}$}     & \colhead{(days)}} 
\startdata
35&2.75$\pm$0.12&135.8$\pm$1.3&2,452,054.64583\\
35&0.43$\pm$0.13&$163.2\pm$8.7&2,452,141.72222\\
86&10.70$\pm$0.41&131.3$\pm$1.1&2,452,830.75347\\
86&12.41$\pm$0.37&15.4$\pm$0.8&2,452,835.78125\\
\enddata
\end{deluxetable}

\clearpage

\begin{deluxetable}{lrrrr} 
\tablewidth{0pt}
\tabletypesize{\scriptsize}
\tablecaption{NIR polarization data of \citet{Jones2003} and equivalent 
$\lambda=7660 {\rm \AA}$ polarization data
in the core and in the filament regions. 
Visible extinction coefficients are from \citet{Dobashi2005}.
\label{IRTOVIS}}
\tablehead{
  \colhead{Star}                       & \colhead{$P_{\rm H} \pm \sigma_{P_{\rm H}}$}              &
  \colhead{$P_{7660} \pm \sigma_{P_{7660}}$}     & \colhead{$A_{\rm V}$ }                        \\
 \colhead{}                       & \colhead{$(\%)$}              &
  \colhead{$(\%)$}     & \colhead{(Mag.)}} 
\startdata
C1&1.9$\pm$0.6&$3.7\pm1.2$&2.34\\
C2&1.4$\pm$0.3&$2.8\pm0.8$&2.34\\
C3&4.2$\pm$1.0&$8.3\pm2.0$&2.60\\
C4&1.7$\pm$0.3&$3.4\pm0.8$&2.34\\
C5&4.0$\pm$0.9&$7.9\pm1.8$&2.60\\
C6&2.7$\pm$0.6&$5.3\pm1.2$&2.34\\
F2&1.9$\pm$0.4&$3.7\pm0.8$&2.19\\
\enddata
\end{deluxetable}
\clearpage

\begin{deluxetable}{lrrrr} 
\tablewidth{0pt}
\tabletypesize{\scriptsize}
\tablecaption{Means and dispersions of polarization and position angles at various wavelengths
and various scales (see figures \ref{mapcore}, \ref{mapfila}, \ref{polmap} and \ref{mapgal} respectively).
\label{MEANS}}
\tablehead{
  \colhead{}                       & \colhead{Core region}              &
  \colhead{ Filament region}              & \colhead{ GF 9 scale }     
& \colhead{Galactic scale }} 
\startdata
$\overline{P}_{7660,\rm V} \pm S_{{P_{7660,\rm V}}}$ ($\%$)&$2.17 \pm 0.93$ \tablenotemark{(a)}&$2.62\pm1.27$ \tablenotemark{(a)}&$2.22 \pm 1.33$ \tablenotemark{(a)}&$1.44\pm1.26$ \tablenotemark{(b)} \\
$\overline{\theta}_{7660,\rm V}  \pm S_{{\theta_{7660,\rm V}}}$ ($^{\circ}$) \tablenotemark{(c)}&$136.0 \pm 12.4$&$126.1\pm13.1$&$127.4\pm25.5$&$72.0\pm43.3$\\
$\overline{P}_{\rm H} \pm S_{{P_{\rm H}}}$ ($\%$)&$2.65 \pm 1.10$&$1.9\pm0.4 $ \tablenotemark{(d)}&&\\
$\overline{\theta}_{\rm H}  \pm S_{{\theta_{\rm H}}}$ ($^{\circ}$) \tablenotemark{(c)}&$171.2\pm5.9$&$137.0 \pm 6.0$ \tablenotemark{(d)}&&\\
$\overline{P}_{\rm FIR} \pm S_{{P_{\rm FIR}}}$ ($\%$)&...&...&&\\
$\overline{\theta}_{\rm FIR}+90^{\circ}  \pm S_{{\theta_{\rm FIR}}}$ ($^{\circ}$)\tablenotemark{(c)} &$\approx 19$&random&&\\
\enddata
\tablenotetext{(a)}{ Our data at $\lambda = 7660 {\rm \AA}$.}
\tablenotetext{(b)}{30$^{\circ} \times 30^{\circ}$ map centered at position ($l_{c}=95^{\circ}, b_{c}=+10^{\circ}$) (see Figure \ref{mapgal}), but in the V band.}
\tablenotetext{(c)}{Position angles are given in the equatorial frame measured east from north.}
\tablenotetext{(d)}{Only one measurement is used at this wavelength (see Figure \ref{mapfila}). }

\end{deluxetable}

\clearpage
\begin{deluxetable}{lrrrrrrrr} 
\tablewidth{0pt}
\tabletypesize{\scriptsize}
\tablecaption{Position angles of cores in GF 9 and offsets with the mean magnetic field position angles as measured in the three different passbands.  
\label{OFFSETS}}
\tablehead{
  \colhead{Core}                       & \colhead{$\alpha (2000)$}              &
  \colhead{$\delta (2000)$}     & \colhead{P.A.\tablenotemark{(a)}}           &
  \colhead{$|\Delta \theta_{7660}|$}  & \colhead{$|\Delta \theta_{\rm H}|$}    &
  \colhead{$|\Delta \theta_{\rm FIR}|$} & \colhead{Location\tablenotemark{(b)}}                       \\ 
  \colhead{name}              &     \colhead{($^{h}$ $^{\rm mn}$ $^{\rm s})$}    & 
  \colhead{($^{\circ}$ ' '')}              &     \colhead{($^{\circ}$)}    & 
  \colhead{($^{\circ}$)}              &     \colhead{($^{\circ}$)}    & 
  \colhead{($^{\circ}$)}    &   \colhead{} } 
\startdata
LM349\tablenotemark{(c)}&20 49 6.9&60 11 48&47&$\approx 72$&$\approx 90$&random&F\\
LM350\tablenotemark{(c)}&20 49 46.5&60 15 40&49&$\approx 70$&$\approx 88$&random&F\\
LM351\tablenotemark{(c)}&20 51 28.0&60 18 33&43&$\approx 91$&$\approx 51$&$\approx 24 $&C\\
LM352\tablenotemark{(c)}&20 53 49.9&60 11 27&...&...&...&...&...\\
L1082C\tablenotemark{(d)}&20 51 27.4&60 19 00&7&$\approx 53 $&$\approx 15 $&$\approx 12 $&C\\
L1082A\tablenotemark{(d)}&20 53 29.6&60 14 41&...&...&...&...&...\\
L1082B\tablenotemark{(d)}&20 53 50.3&60 09 47&48&?&?&?&...\\
\enddata
\tablenotetext{(a)}{East of north.}
\tablenotetext{(b)}{F: filament region, C: core region.}
\tablenotetext{(c)}{\citet{Lee1999}.}
\tablenotetext{(d)}{\citet{Benson1989}.}
\end{deluxetable}

\clearpage
\begin{deluxetable}{ll} 
\tablewidth{0pt}
\tabletypesize{\scriptsize}
\tablecaption{Magnetic field strength in CS cores. 
\label{STRENGTHS}}
\tablehead{
  \colhead{Parameter}                       & \colhead{GF9 Core}     \\
    \colhead{}                       & \colhead{\tablenotemark{(a)}} }
\startdata
 $n$ (H$_{2}$ cm$^{-3}$)\tablenotemark{(b)}           &15000 $\pm$ 3000\\
 $N$ (H$_{2}$ cm$^{-2}$)\tablenotemark{(b)}       &9.9 $10^{21}$\\
$R (pc)$\tablenotemark{(b)}                           &0.16\\
$M $ $(M_{\bigodot})$\tablenotemark{(b)}             &$15\pm3$\\
$T_{K} (K)$\tablenotemark{(b)}                       &$7.5 \pm 0.5$\\
$\Delta \rm V$ ($km/s$)\tablenotemark{(b)}           &0.88$\pm$ 0.20\\
$\delta \phi$ $(^{\circ})$                                &$5.9$\\
$B_{\rm pos}$ ($\mu G$)                                &170$\pm$56\\
$\lambda_{\rm obs}$\tablenotemark{(c)}                 &$\approx$ 0.44 \\
$\lambda_{\rm c}$                                      & 0.35   \\
\enddata
\tablenotetext{(a)}{Only IR data are used.}
\tablenotetext{(b)}{\citet{Ciardi2000}.}
\tablenotetext{(c)}{$\lambda_{\rm obs} = 7.6 \times 10^{-21} \frac{N_{\rm H_{2}}}{B_{\rm pos}}$, \citet{Crutcher2004}.}  
\end{deluxetable}

\clearpage
\vspace{11.5cm}

\clearpage

\clearpage

\clearpage

\clearpage

\clearpage
\begin{figure}
\epsscale{0.8}
\plotone{f6.eps}
\caption{ 
Polarization map around and into 
the core (C) region observed by \citet{Ciardi1998}, \citet{Ciardi2000}.
Bold vectors represent visible data. Thin vectors are IR data from \citet{Jones2003} with 
$P=1.5 \%$ for clarity. Vectors 
with intermediate size are ISO FIR-90$^{\circ}$ rotated data from \citet{Clemens1999} with 
$P=2.5 \%$ for clarity. Only FIR data for which the position angle is approximately the same
when using the KLC intrumental polarization and the C-off instrumental polarization are shown.     
\label{mapcore}}
\end{figure}

\clearpage
\begin{figure}
\epsscale{0.8}
\plotone{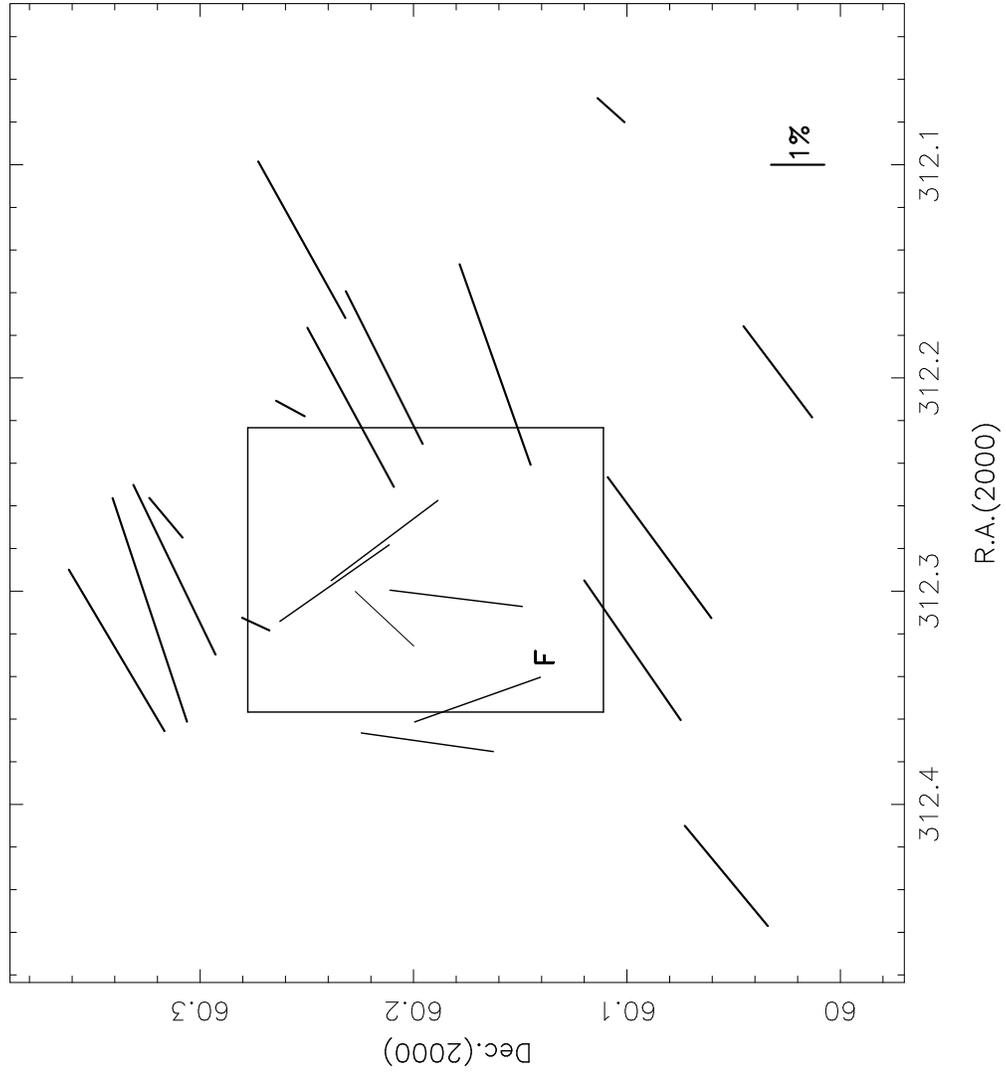}
\caption{ 
Polarization map in the filament region.
Same as in Figure \ref{mapcore}, but around and into the filament (F) 
region observed by \citet{Ciardi1998}, \citet{Ciardi2000}. 
\label{mapfila}}
\end{figure}

\clearpage
\begin{figure} 
\plotone{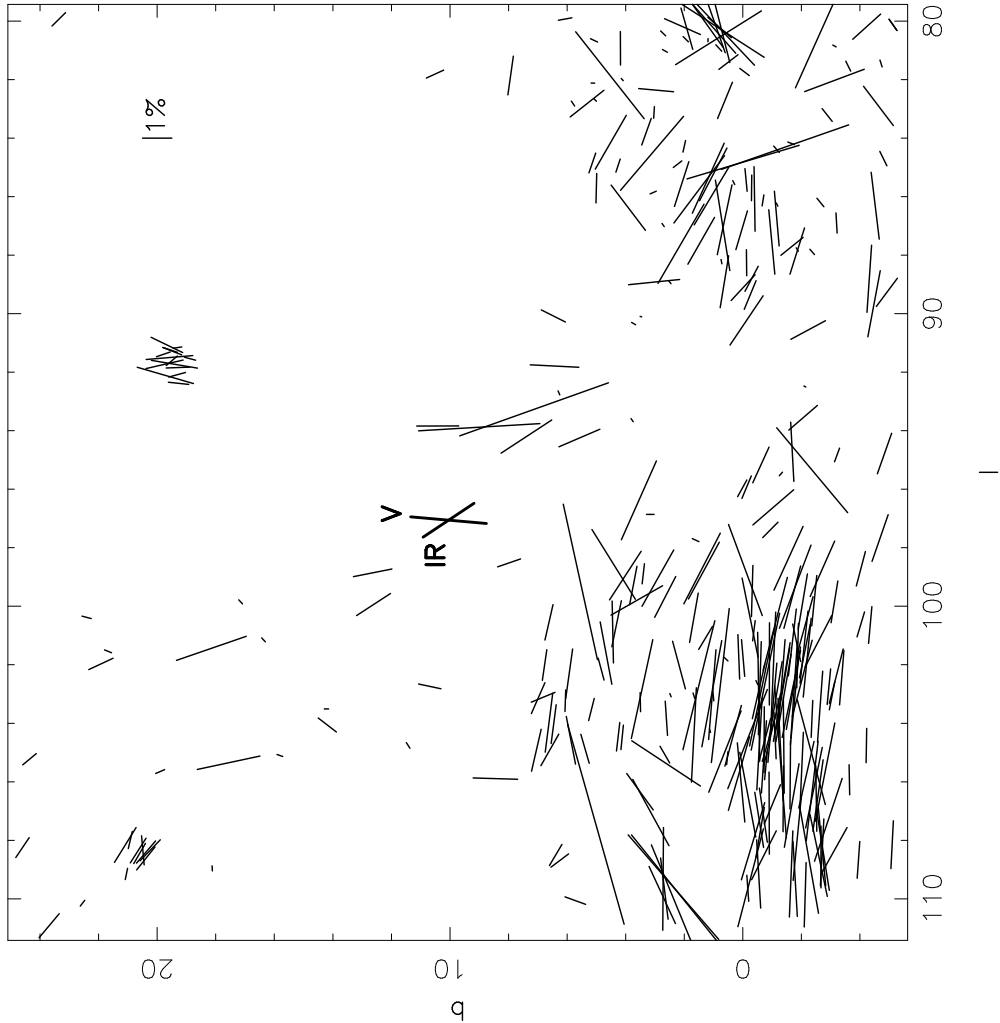}
\epsscale{0.8}
\caption{
Galactic polarization map around GF 9.
Visible data from the Heiles catalog in a $30^{\circ} \times 30^{\circ}$ box centered 
at position $(l_{c}=95^{\circ}, b_{c}=+10^{\circ})$ are shown.
The IR symbol denotes the mean IR polarization of data taken in the GF9 core region 
by \citet{Jones2003}.
The V symbol shows the mean visible polarization from our data in the vicinity of the core 
region.
\label{mapgal}}
\end{figure}

\end{document}